# High burden of private mutations due to explosive human population growth and purifying selection

**Feng Gao[1], Alon Keinan[1,§]**

[1]Department of Biological Statistics and Computational Biology, Cornell University, Ithaca, NY 14853, USA

[§]Corresponding author

Email Addresses:

    FG: fg237@cornell.edu

    AK: ak735@cornell.edu

# Abstract

**Background**

Recent studies have shown that human populations have experienced a complex demographic history, including a recent epoch of rapid population growth that led to an excess in the proportion of rare genetic variants in humans today. This excess can impact the burden of private mutations for each individual, defined here as the proportion of heterozygous variants in each newly sequenced individual that are novel compared to another large sample of sequenced individuals.

**Results**

We calculated the burden of private mutations predicted by different demographic models, and compared with empirical estimates based on data from the NHLBI Exome Sequencing Project and data from the Neutral Regions (NR) dataset. We observed a significant excess in the proportion of private mutations in the empirical data compared with models of demographic history without a recent epoch of population growth. Incorporating recent growth into the model provides a much improved fit to empirical observations. This phenomenon becomes more marked for larger sample sizes, e.g. extrapolating to larger sample sizes, after sequencing 10,000 individuals from the same population with perfect accuracy, still about 1 in 400 heterozygous sites (or about 6,000 variants) at the 10,001$^{st}$ individual are predicted to be novel, 18-times as predicted in the absence of recent population growth. The proportion of private mutations is additionally increased by purifying selection, which differentially affect mutations of different functional annotations.

**Conclusions**

The burden of private mutations for each individual, which are singletons (i.e. appearing in a single copy) in a larger sample that includes this individual, is predicted to be greatly increased by recent population growth, as well as by purifying selection. Comparison with empirical data supports that European populations have experienced recent rapid population growth, consistent with previous studies. Moreover, this comparison supports that purifying selection has had a more pronounced effect on variants that are predicted to be functional. These results have important implications to the design and analysis of sequencing-based association studies of complex human disease as they pertain to private and very rare variants.

# Keywords

# Background

Many recent studies that sequenced large numbers of individuals have shown that human populations have experienced a complex demographic history, including a recent epoch of rapid growth in effective population size, although estimates have varied greatly among studies [1-7]. The growth of European population has recently been estimated to be exponential with a rate of 2-5% per-generation increase in population size [1,3,7]. This recent growth has resulted in an excess of rare single nucleotide variants (SNVs), commonly defined as those with a minor allele (the less common of the two alleles) frequency (MAF) less than 0.5% (or 1%) in a sample of individuals from the same population [e.g. 5,8]. The proportion of singletons (SNVs with only one copy in the entire sample) is especially elevated due to this recent rapid growth [1,3,5,7,9]. Consequently, the corresponding site frequency spectrum (SFS), a summary statistic that indicates the proportion of variants of each possible allele count in the sample (e.g. Figure 1) is skewed towards lower allele counts.

A predicted consequence of the skew in the SFS due to population growth is an increase in the burden of private mutations for each individual. We recently defined this quantity as the proportion of heterozygous positions in each newly sequenced individual that are novel, i.e., completely absent from a previously sequenced sample from the same population [9]. This recent paper observed this burden to be higher in samples from populations of European and East Asian descent than is predicted by previously estimated demographic models that do not include an epoch of recent population growth [9]. However, empirical estimates in that paper were based on a small sample size of less than 100 individuals, while the contribution of recent rapid growth is expected to be more pronounced for larger sample sizes [1-6,9].

Here, we set out to (1) empirically estimate the burden of private mutations from large samples of individuals of European ancestry, (2) compare these estimates with predictions of previously proposed demographic models with and without a recent epoch of exponential growth [3,10], and (3) contrast SNVs of different functions that are expected to have undergone different selective effects. As purifying, negative selection on deleterious SNVs skews the SFS towards rare variants [1,5,11,12,13], it can interact with the effect of recent population growth in increasing the burden of private SNVs, and differently so for different functional categories. With the rapidly decreasing cost of sequencing, more and more high-quality sequencing data sets of large sample sizes and improved accuracy of detecting rare variants become available. This provides an excellent opportunity for a more accurate study of the burden of private mutations. In this paper, we considered two such sequencing data sets of samples from European populations: the Neutral Regions (NR) data set of putatively neutral regions [3] and that from the NHLBI Exome Sequencing Project (ESP) [1,7].

## Results and Discussion

In all analyses, we contrast three different demographic models and the fit of their predictions to the NR data set [3] and to 7 functional categories of the ESP data set [1,7]. The three demographic models are (1) a population that has been of constant population size throughout history, (2) a model of European history that includes two population bottlenecks [10], and (3) a model of European history with two bottlenecks, a recent change in population size, followed by a recent epoch of rapid population growth [3] (Model II therein).

**Comparison of site frequency spectra**

As the burden of private mutations is a function of the site frequency spectrum, we first contrasted the site frequency spectra between three demographic models, the NR data [3], and the ESP data [1,7] (Figure 1). In order to allow comparison of the data sets with different sample sizes, as well as account for missing genotype calls for each SNV, we probabilistically subsampled all data to a sample size of 900 haploid chromosomes (Methods). The proportion of singletons from demographic models (1) and (2) is greatly lower than that in the observed data and that predicted by model (3) that incorporates recent growth (Figure 1). Among the categories of the ESP data, categories that are expected to be more functional show a higher proportion of singletons, e.g. intronic, intergenic, synonymous, and UTR SNVs have a significantly lower proportion than non-synonymous, nonsense, and splice SNVs (Figure 1), which is expected by the latter being more often deleterious. The proportion of singletons in the SNVs from the NR data is lower than all categories of SNVs from ESP, which is consistent with the former being designed such that variants are very far from genes and putatively neutral [3], while the latter consists of variants in and near protein-coding genes [1,7], which are expected to more often be targeted by purifying selection. Another factor that can contribute to this difference between the NR and ESP datasets is that the former aimed to capture a sample of homogenous ancestry, which corresponds to North-Western European ancestry [3], while the latter consists of a broad sample of European Americans that exhibits a higher level of population structure [1,7]. Increased population structure can lead to an increase in the proportion of rare variants since some of these can postdate the split of the population captured by the different ancestries [3].

These site frequency spectra recapitulate results from the corresponding previous studies [1,3,7] and were presented here to motivate the study of the burden of private mutations in the following. The differences in the SFS between functional categories of the ESP data were also shown by the original studies [1,7].

**Comparison of the burden of private mutations**

The predicted burden of private mutations for each individual from all demographic models and the empirical burden observed in the different data sets and functional categories are presented in Figure 2. Across all sample sizes, the burden of private mutations from empirical data is significantly higher than that predicted by demographic models without growth. For example, based on the results of the NR data, when 100 individuals have been sequenced, we estimated that about 1.4% out of all heterozygous sites in the 101$^{st}$ sequenced individual are novel, that is specific to the 101$^{st}$ individual and completely absent from the first set of 100 individuals. While models (1) and (2) predict only 1% in this scenario, model (3) is consistent with this estimate in the NR data.

For all demographic models and observed data, as more individuals are sequenced, the burden of private mutations decreases (Figure 2), because increasing sample size makes it more probable that a variant has already been discovered [9]. However, the effect of recent growth itself on the burden of private mutations is much more pronounced with increasing sample size. For example, for the NR data, when 492 individuals are sequenced, the estimated burden of mutation from the 493$^{rd}$ sequenced individual is about 0.76% (Table 1). The estimations from models (1) and (2) are only 0.20% and 0.26%, respectively, about a third of empirical data, while model (3) matches the data well. We note that this percentage varies greatly across individuals with the relatively small number of SNVs in the NR data (Table 2). When extrapolating the models to consider a scenario in which 10,000 individuals are sequenced, model (3) predicts the burden of mutations of the 10,001$^{st}$ individual to be 0.24% (Table 1), 24-times and 18-times that from models without recent growth that predict 0.010% and 0.013% based on models (1) and (2), respectively (Table 1). This corresponds to almost 1 of 400 heterozygous positions, which is equivalent to about 6,000 variants. This estimate is two orders of magnitude larger than the expected number of de novo mutations of each individual [e.g. 14]. Hence, we predict that thousands of novel variants will be discovered in each newly sequenced genome even after tens of thousands of genomes from exactly the same population have already been sequenced with perfect accuracy.

Another important observation is that the burden of private mutations for each individual calculated from all seven categories of the ESP data is consistently higher than that from the NR data for all sample sizes (Figure 2). This is consistent with the observation that the SFS's of the ESP data are more left-skewed than those of the NR data, which is consistent with decreased effect of purifying selection population structure on the latter. Comparing the different ESP categories, splice and non-sense SNVs, which are expected to most often be deleterious, have the largest burden of private mutations across all sample sizes. Similarly, the burden of all functional categories is ordered by common expectations as to how likely such mutations are expected to be functional. The burden of private mutations captures a unique summary of the SFS that more clearly shows the effect of purifying selection. For example, when $n = 492$, the proportion of singletons is 46.2% for the ESP intergenic SNVs and 74.8% for the ESP splice SNVs, which is 1.6-fold. In comparison, the burden of private mutations for splice SNVs is about 9.7-fold of that for intergenic SNVs. This difference is even more pronounced when the sample size is larger, with 12.7-fold different when $n = 4299$ (Figure 2).

## Conclusions

Recent whole-genome sequencing data sets show that the proportion of rare variants in large samples, especially singletons, is significantly elevated compared with the prediction from the standard coalescent theory that assumes a constant population size and from previous demographic models without recent growth [1,3,7,9]. Recent demographic modeling studies predict that humans have experienced a recent and rapid population growth, which explains an increased proportion of singletons and other rare variants [1-6]. In this paper, we examined the burden of private mutations for each individual, a statistic that reflects the relationship between the relative proportions of singletons and more common variants contained in a sample, with three demographic models and two data sets under different sample sizes. We found that the burden of private mutations calculated from empirical data and estimated from demographic models with a recent growth is significantly higher than that estimated from models without recent growth across all sample sizes. The discrepancy is predicted to be much more pronounced if the number of sequenced individuals becomes larger. We showed that this finding is consistent with a recent epoch of population growth. Moreover, we found that the SNVs that are affected by stronger purifying selection will generally have larger burden of private mutations compared with more selectively neutral SNVs, since they will have a higher proportion of singletons.

The proportion of private mutations that we consider translated to the number of novel variants expected to be ascertained with each newly sequenced genome. Hence, our results have implications to sequencing-based association studies of complex human diseases and other sequencing studies. For instance, we predict that even after 10,000 individuals from the exact same European population have been perfectly sequenced, still 1 in 400 heterozygous sites will be novel in each newly sequenced genome, which corresponds to discovering about 6,000 new variants. This large expectation is due to the effect of the recent rapid growth of European populations, which leads to this number being at least 18-fold that predicted in the absence of such growth. Hence, careful consideration must be given to private mutations in the design and analysis of sequencing-based association studies and in quantifying the role played by rare variants in complex human disease [15-19].

## Methods

**Datasets**

Two data sets were used in this study. The NR data contains the genotypes of 493 European individuals with high homogeneity on relatively neutral SNVs of 15 genetic regions [3]. For quality purposes, all SNVs with less than 900 successful genotype counts were filtered from the analysis. The remaining 1,746 SNVs constitute 95% of all variants [3]. The summarized data of 4,300 European individuals from NHLBI Exome Sequencing Project records the minor allele count and major allele count of each SNV identified in 15,585 genes on all chromosomes (including chromosome X and Y) [1,7]. In this analysis, we combined all of the autosomal SNVs according to the 7 categories: intergenic, intron, missense, nonsense, splice, synonymous and UTR. For quality purpose, SNVs are filtered if the average read depth is less than or equal to 20 or the successful genotype counts are less than 8,170 (95%).

**Subsampling approach**

In order to compare the SFS of data with different sample sizes (including the different sample sizes across the SNVs caused by unsuccessful genotype counts in the same data set), all the observed data were subsampled to 900 chromosomes. Following the strategy used in [10], for a SNV with $j$ minor alleles out of $n$ successful genotype counts, the probability that it is of $x$ minor alleles when subsampled to $m$ chromosomes is

$$P(x \text{ of } m) = \frac{\binom{j}{x}\binom{n-j}{m-x}}{\binom{n}{m}} + \frac{\binom{j}{m-x}\binom{n-j}{x}}{\binom{n}{m}}$$

where $x = 0, 1, 2, \ldots, \left[\frac{m}{2}\right]$ and $\binom{a}{b} := 0$ if $a < b$.

**Expected SFS and the burden of private mutations for demographic models**

The SFS of the three demographic models were calculated using exact computation [20] instead of simulations.

For a demographic model with constant population size, the burden of private mutations can be derived under standard coalescent theory [21]. For constant population size, the expected number of singletons of a folded SFS for a sample of ($n$ + 1) diploid individuals is

$$E[\eta_1] = \theta \left(1 + \frac{1}{2n+1}\right)$$

where $\theta = 4N\mu$. The expected number of singletons that belong to one individual is

$$E[s] = \frac{1}{n+1} E[\eta_1] = \frac{2\theta}{2n+1}$$

The expected number of heterozygote sites for the pair of sequences from one individual $E[h] = \theta$. Thus the expected burden of private mutations is

$$E[\alpha] = \frac{E[s]}{E[h]} = \frac{2}{2n+1}$$

For variable population size, the general solution is

$$E[\alpha] = \frac{1}{n+1} \frac{E[T_{2n+2,1}] + E[T_{2n+2,2n+1}]}{E[T_{2,1}]}$$

where $T_{p,q}$ stands for the total length of all branches in the coalescent tree which have exactly $q$ descents out of the total number of descents $p$. The branch lengths are calculated by exact computation [20].

**Computation of the burden of private mutations using data sets and simulations**

For the NR data, for each of the 493 individuals, the burden of private mutations $\alpha$ is directly calculated by the proportion of heterozygote sites which contain singletons using the individual genotypes. Missing genotypes were abandoned. The mean and standard deviation of $\alpha$ for this sample were then calculated by

$$\bar{\alpha} = \frac{1}{n}\sum_{i=1}^{n}\alpha_i, \, s(\alpha) = \sqrt{\frac{\sum_{i=1}^{n}(\alpha_i - \bar{\alpha})^2}{n-1}}$$

where $n$ is the sample size and equals 493 here.

For ESP data and demographic models, as the individual genotypes were not available, sequences were simulated by distributing the minor alleles of each SNV to individuals randomly and independently. Unsuccessful genotype calls (missing genotypes) were also distributed randomly to the individuals but were distributed in pairs. In other words, the genotypes of each individual at each site either were both existent or both missing. Then $\alpha$ was calculated using these simulated sequences in the same way as for the NR data.

For the demographic histories from which we can only get the SFS, a similar method is applied. Namely we simulated a certain number of SNVs according to the SFS and randomly assigned the minor alleles into individual sequences. The simulated sequences were paired randomly to form the sequences of an individual and $\alpha$ for each individual was then calculated.

To calculate $\alpha$ for a smaller sample size $m$, $m$ individuals were randomly chosen from the original $n$ individuals and $\alpha$ was calculated using the genotypes from these $m$ individuals with the previously stated approach.

To study the effects of limited sites, a bootstrap approach was applied. Specifically, we resampled individual SNPs with replacement for 1,000 times. For each bootstrap, we calculated the average $\alpha$ ($\alpha_{b,i}$) across all individuals and these 1,000 averages were used to calculate the mean and standard deviation of the bootstrap, the latter of which is an estimate of the standard error of the sample:

$$\bar{\alpha}_b = \frac{1}{n_b}\sum_{i=1}^{n_b} \alpha_{b,i}, \; s_b(\alpha) = \sqrt{\frac{\sum_{i=1}^{n_b}(\alpha_{b,i}-\bar{\alpha}_b)^2}{n_b-1}}$$

where $n_b$ is the number of bootstraps and equals 1,000 here.

## Competing interests

The authors declare that they have no competing interests.

## Authors' contributions

FG carried out all analyses. AK conceived the experiments and provided materials. Both authors co-wrote the manuscript and approved its final version.

## Acknowledgements

We thank Diana Chang for helpful comments on previous versions of this manuscript. This work was supported in part by National Institutes of Health Grant R01HG006849.

A.K. was also supported by The Ellison Medical Foundation and the Edward Mallinckrodt, Jr. Foundation.## References

1. Tennessen JA, Bigham AW, O'Connor TD, Fu W, Kenny EE, Gravel S, McGee S, Do R, Liu X, Jun G, Kang HM, Jordan D, Leal SM, Gabriel S, Rieder MJ, Abecasis G, Altshuler D, Nickerson DA, Boerwinkle E, Sunyaev S, Bustamante CD, Bamshad MJ, Akey JM; Broad GO; Seattle GO; NHLBI Exome Sequencing Project: **Evolution and functional impact of rare coding variation from deep sequencing of human exomes.** *Science* 2012, **337**(6090):64-69.

2. Gravel S, Henn BM, Gutenkunst RN, Indap AR, Marth GT, Clark AG, Yu F, Gibbs RA; 1000 Genomes Project, Bustamante CD: **Demographic history and rare allele sharing among human populations.** *Proc Natl Acad Sci U S A* 2011, **108**(29):11983-11988.

3. Gazave E, Ma L, Chang D, Coventry A, Gao F, Muzny D, Boerwinkle E, Gibbs R, Sing CF, Clark AG, Keinan A: **Neutral genomic regions refine models of recent rapid human population growth.** *Proc Natl Acad Sci U S A* 2014, **111**(2):757-762.

4. Gutenkunst RN, Hernandez RD, Williamson SH, Bustamante CD: **Inferring the joint demographic history of multiple populations from multidimensional SNP frequency data.** *PLoS Genet* 2009, **5**(10):e1000695.

5. Nelson MR, Wegmann D, Ehm MG, Kessner D, St Jean P, Verzilli C, Shen J, Tang Z, Bacanu SA, Fraser D, Warren L, Aponte J, Zawistowski M, Liu X, Zhang H, Zhang Y, Li
- 16 -

# Figures

**Figure 1. SFS of demographic models and data with a sample size of 900**

The SFS for 3 demographic models, the NR data and 7 categories of the ESP data. To adjust for the different sample sizes in the two datasets, probabilistic subsampling was applied to make all sample sizes equal to 900 chromosomes. Only the first 10 minor allele count categories are shown. For each minor allele count, from left to right: constant population size, European history with 2 bottlenecks but no growth [10], European history with recent growth (Model II in [3]), the NR data, intergenic SNVs of the ESP data, intron SNVs of the ESP data, synonymous SNVs of the ESP data, UTR SNVs of the ESP data, missense SNVs of the ESP data, nonsense SNVs of the ESP data and splice SNVs of the ESP data.

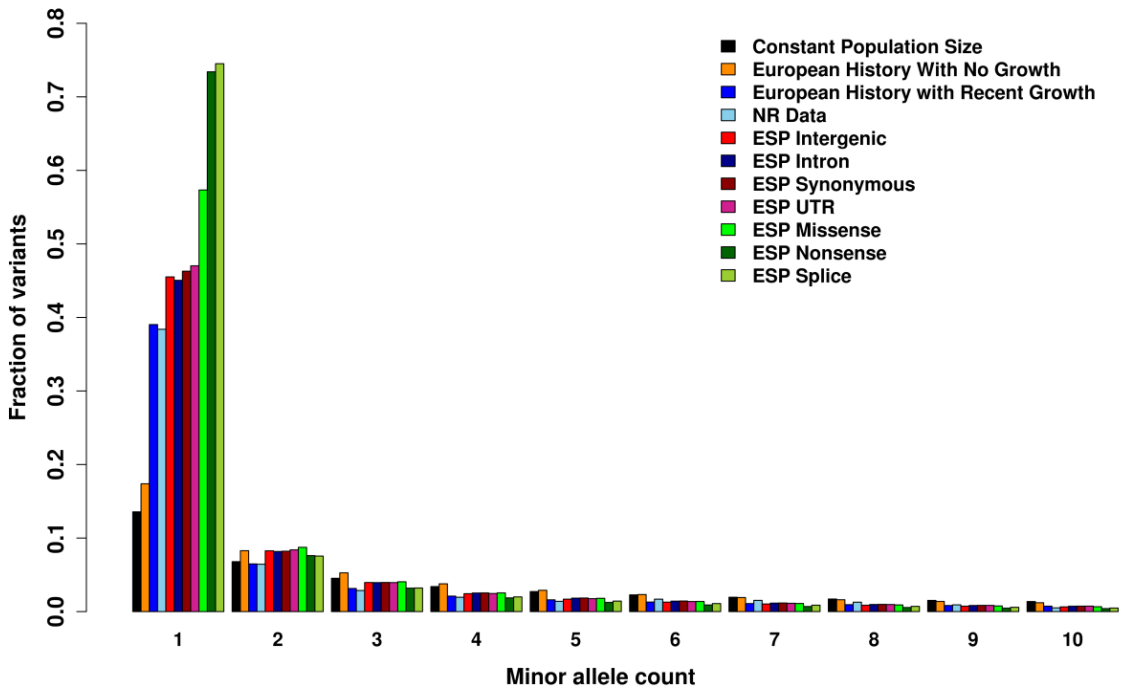

**Figure 2. The burden of private mutations of demographic models and empirical data**

The burden of private mutations for the same demographic models and empirical data as in Figure 1, using the same colors. This quantity corresponds to the percentage out of all heterozygous sites in a newly sequenced genome that are novel after *n* genomes have already been sequenced. Results are presented for *n* = 100, *n* = 492, *n* = 1000, *n* = 4299 and *n* = 10000. The value of 492 and 4299 are dictated by the sample size of the NR and ESP datasets, respectively. For empirical data, mean percentage across individuals is presented, together with error bars that denote ± one standard error across SNVs, estimated via bootstrapping (Methods). Double-slashes around a value of 0 on the *x*-axis represent instances where data for that sample size is not available in the respective datasets. Note that the range above 5% on the *y*-axis is rescaled. The corresponding values in this figure are shown in Table 1.

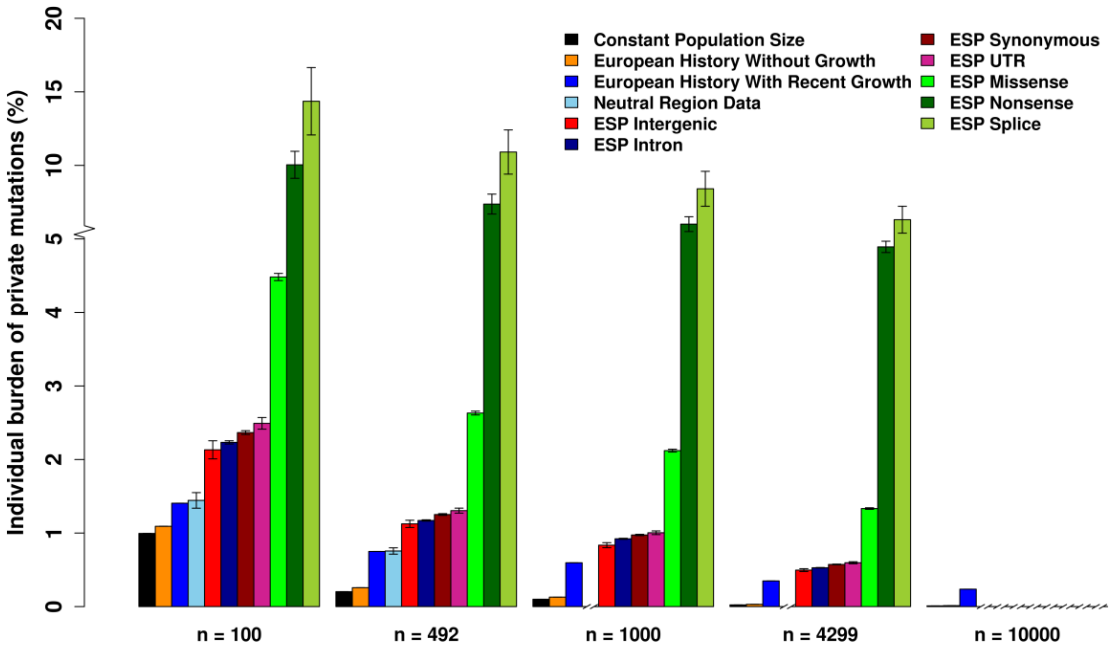

# Tables

**Table 1. Estimated mean and standard error of percentage of private mutations for each individual**

The burden of private mutations for $n = 100$, $n = 492$, $n = 1000$, $n = 4299$ and $n = 10000$, the corresponding values for Figure 2. This table was provided for completeness. The number in parenthesis denotes the standard error across SNVs estimated via bootstrap (Methods). NA indicates that the data for that sample size is not available in the respective datasets.

| Group | $n = 100$ | $n = 492$ | $n = 1000$ | $n = 4299$ | $n = 10000$ |
|---|---|---|---|---|---|
| Constant Population Size Model | 0.995% | 0.203% | 0.100% | 0.023% | 0.010% |
| European History with Two Bottlenecks | 1.092% | 0.257% | 0.129% | 0.031% | 0.013% |
| European History with Recent Growth | 1.406% | 0.750% | 0.596% | 0.349% | 0.237% |
| NR Data | 1.444% (0.106%) | 0.756% (0.043%) | NA | NA | NA |
| ESP Intergenic | 2.132% (0.123%) | 1.125% (0.049%) | 0.835% (0.034%) | 0.496% (0.019%) | NA |
| ESP Intron | 2.233% (0.022%) | 1.171% (0.009%) | 0.922% (0.007%) | 0.528% (0.004%) | NA |
| ESP Synonymous | 2.366% (0.026%) | 1.252% (0.012%) | 0.974% (0.009%) | 0.573% (0.005%) | NA |
| ESP UTR | 2.492% (0.079%) | 1.305% (0.034%) | 1.004% (0.025%) | 0.596% (0.014%) | NA |
| ESP Missense | 4.482% (0.049%) | 2.632% (0.026%) | 2.121% (0.019%) | 1.333% (0.011%) | NA |
| ESP Nonsense | 10.04% (0.92%) | 7.37% (0.68%) | 6.00% (0.50%) | 4.46% (0.38%) | NA |
| ESP Splice | 14.36% (2.29%) | 10.91% (1.50%) | 8.41% (1.19%) | 6.31% (0.91%) | NA |

**Table 2. The mean and standard deviation of the burden of private mutations across individuals**

The burden of private mutations and the standard deviation of the sample for three demographic models and the NR data. The results correspond to $n = 492$, the sample size of the NR data less one. The real genotypes were used for the calculation of the NR data. For the three demographic models without real genotypes, sequences were simulated with the same number of SNVs as in the NR data (Methods). The number in parenthesis denotes the standard deviation of the sample. These large standard deviations suggest significant variations in percentage of private mutations across individuals with the relatively small number of SNVs in the NR dataset.

| Group | The Burden of Private Mutations |
|---|---|
| Constant Population Size Model | 0.208% (0.299%) |
| European History with Two Bottlenecks | 0.276% (0.352%) |
| European History with Recent Growth | 0.736% (0.614%) |
| NR Data | 0.758% (0.852%) |